\documentclass[a4paper,preprint,showpacs,amssymb,superscriptaddress]{revtex4}
\usepackage{graphicx}

\newcommand{\be}{\begin{equation}}
\newcommand{\ee}{\end{equation}}
\newcommand{\ben}{\begin{eqnarray}}
\newcommand{\een}{\end{eqnarray}}

\newcommand{\tr}{{\mathrm{Tr}}}
\begin{document}
\draft
\title{Jensen-Shannon divergence as a measure of distinguishability between mixed quantum states}
\author{A.P. Majtey}
\address{Facultad de Matem\'atica, Astronom\'\i a y F\'\i sica \\
Universidad Nacional de C\'ordoba \\ Ciudad Universitaria, 5000
C\'ordoba, Argentina \\ CONICET}
\author{P.W. Lamberti}
\address{Facultad de Matem\'atica, Astronom\'\i a y F\'\i sica \\
Universidad Nacional de C\'ordoba \\ Ciudad Universitaria, 5000
C\'ordoba, Argentina \\ CONICET}
\author{D.P. Prato}
\address{Facultad de Matem\'atica, Astronom\'\i a y F\'\i sica \\
Universidad Nacional de C\'ordoba \\ Ciudad Universitaria, 5000
C\'ordoba, Argentina}

\date{\today}

\begin{abstract}
We discuss an alternative to relative entropy as a measure of
distance between mixed quantum states. The proposed quantity is an
extension to the realm of quantum theory of the Jensen-Shannon
divergence (JSD) between probability distributions. The JSD has
several interesting properties. It arises in information theory
and, unlike the Kullback-Leibler divergence, it is symmetric,
always well defined and bounded. We show that the quantum JSD
(QJSD) shares with the relative entropy most of the physically
relevant properties, in particular those required for a ``good''
quantum distinguishability  measure. We relate it to other known
quantum distances and we suggest possible applications in the
field of the quantum information theory.
\end{abstract}


\maketitle

\section{Introduction}
Distance measures play a central role in quantum theory, in
particular in the context of quantum computation and quantum
information. On one side they allow us to establish precisely the
problem of quantum state discrimination \cite{Nielsen}; on the
other they are associated to the definition of the degree of
entanglement, just to mention two very significant subjects
\cite{Vedral2}.

Several distances between quantum states have been introduced.
Many of them were defined as distances (divergences) between
probability distributions and then extended as distances between
quantum states. This is the case of the Hellinger distance, which,
for two (discrete or continuous) probability distributions $P(x)$
and $Q(x)$ reads
\begin{equation}
H(P,Q) = \sum_x [\sqrt{P(x)}- \sqrt{Q(x)}]^2
\end{equation}
and its quantum mechanical version is expressed as
\begin{equation}
H(\rho\|\sigma) = \tr(\sqrt{\rho} - \sqrt{\sigma})^2
\end{equation}
where $\rho$ and $\sigma$ are two density operators and $\tr$
stands for the trace operator. Analogously the Kullback-Leibler
divergence $ S(P,Q)= \sum_x P(x) \log  \{\frac{P(x)}{Q(x)}\}$ and
the Kolmogorov distance, $K(P,Q)=\frac{1}{2} \sum_x |P(x)-Q(x)|$
are extended to the realm of quantum mechanics. In the first case
the resulting distance is known as the relative entropy
\cite{Lindblad}, \cite{Wehrl}; in the second one the quantum
counterpart is given by \cite{Fuchs}
\[
K(\rho\|\sigma)= \frac{1}{2} \tr|\rho-\sigma|.
\]
Alternatives to these quantum distances have been recently
proposed \cite{Luo},\cite{Lee},\cite{Abe}.

As we mentioned previously a basic issue in quantum information
theory is to distinguish two quantum states by quantum
measurement. In a seminal paper Wootters investigated this problem
and introduced a ``distinguishability distance'' between two pure
states \cite{Wootters}. He defines this distance as the number of
distinguishable intermediate states between the two states.
Braunstein and Caves extended this distance to density operators
for mixed states \cite{Caves} and framed it in a general geometric
formulation of quantum states space \cite{BC}.

In a recent paper we revised the Wootters's distinguishability
distance in terms of the Jensen-Shannon divergence (JSD)
\cite{nosotros}. The JSD between the probability distributions
$P(x)$ and $Q(x)$ is defined by:
\begin{eqnarray}
JS(P,Q)& \equiv & S(P, \frac{P+Q}{2}) + S(Q, \frac{P+Q}{2})
\nonumber \\
& = & H_S(\frac{P+Q}{2}) - \frac{1}{2} H_S(P) - \frac{1}{2} H_S(Q)
\label{basica}
\end{eqnarray}
where $S(P,Q)$ is the Kullback-Leibler divergence and $H_S(P) =
-\sum_x P(x) \log P(x)$ is the Shannon entropy. This quantity was
introduced by C. Rao \cite{Rao} and J. Lin \cite{Lin} as a
symmetrized version of the Kullback-Leibler divergence and has
been recently applied to the analysis of symbolic sequences as
well as to other problems of interest in statistical physics
\cite{Grosse}. It originates in information theory, it is always
well defined and bounded and its square root is a true metric for
the probability distributions space (i.e., its square root is
symmetric, null only when the probability distributions coincide
and it verifies the triangle inequality)\cite{Endres}. It has
several interesting interpretations; for example in statistical
inference theory it gives both the lower and upper bounds to
Bayes' probability error \cite{Lin}. In the framework of
information theory the JSD can be related to mutual information
\cite{Grosse}.

>From some results of harmonic analysis it is possible to show that
the metric space $(X_N^+, \sqrt{JS})$, where $X_N^+$ denotes the
discrete probability distributions space, can be isometrically
mapped into a subset of a Hilbert space \cite{Schoe}, \cite{Berg}.
This result establishes a connection between information theory
and differential geometry, which, we think, could have interesting
consequences in the realm of quantum information theory.

In reference \cite{Lin}, J. Lin proposed a generalization of Eq.
(\ref{basica}) as a distance for several probability
distributions. In fact, let $P_1(x),...,P_N(x)$ be a set of
probability distributions and let $\pi_1,...,\pi_N$ be a
collection of non-negative numbers such that $\sum_i \pi_i =1$.
Then the JSD of the probability distributions $P_i(x), i=1...N$ is
defined by
\begin{equation}
JS^{(\pi_1,...,\pi_N)}(P_1,...,P_N) = H_S(\sum_{i=1}^N \pi_i P_i)
- \sum_{i=1}^N \pi_i H_S(P_i) \label{varios}
\end{equation}
A remarkable feature of this generalized JSD is that it is
possible to assign different weights to the distributions $P_i$.
This might be of importance in the study of quantum statistical
inference problems \cite{Brady}.

In this work we extend the JSD into the context of quantum theory
as a distance measure between mixed quantum states. We show that
it shares with the relative entropy many of the most physically
relevant properties. We also relate it to other commonly used
distances as well as we suggest its applicability as a good
measure of entanglement and fidelity. We see that, when expression
(\ref{varios}) is extended to quantum theory, it can be
interpreted as the upper bound for quantum accessible information.

The structure of the paper is as follows. In the following section
we review the basic properties of the relative entropy, mainly to
guide us in our investigation of the properties satisfied by the
QJSD. In section III we introduce the QJSD and enumerate its basic
properties. Finally we discuss possible applications of the QJSD
in the field of quantum information theory.

\textit{Notation remark}: We will use the following notation:
$D(\; ,\;)$ denotes a distance defined between probability
distributions; $D(\;\|\;)$ denotes the corresponding distance
between density operators.

\section{Relative entropy}

Let $\cal{H}$ be the Hilbert space associated to a quantum system
and let $\cal{S}$ be the set of density operators describing the
states of the system; i.e. $\cal{S}$ is the set of self-adjoint,
positive and trace unity operators.

The relative entropy of an operator $\rho$, with respect to an
operator $\sigma$, both belonging to $\cal{S}$, is given by
\begin{equation}
S(\rho\|\sigma)=\tr[\rho(\log\rho-\log\sigma)]
\end{equation}
where $\log$ stands for logarithm in base two. $S(\rho\|\sigma)$
is nonnegative and vanishes if and only if $\rho = \sigma$. It is
nonsymmetric and unbounded. In particular, the relative entropy is
well defined only when the support of $\sigma$ is equal to or
larger than that of $\rho$. Otherwise, it is defined to be
$+\infty$ \cite{Lindblad}; (the support of an operator is the
subspace spanned by the eigenvectors of the operator with non-zero
eigenvalues). This is a very restrictive requirement which is
violated in some physically relevant situations, for example when
$\sigma$ is a pure reference state \cite{Abe}.

Now we list the fundamental properties of the relative entropy.
For the proofs and a detailed discussion see references
\cite{Wehrl} and \cite{Vedral}:
\begin{enumerate}
\item $S$ is invariant under unitary transformations, that is
\begin{equation}
S(U\rho U^\dag\|U \sigma U^\dag) = S(\rho\|\sigma)
\end{equation}
for any unitary operator $U$. This is a quite natural property to
be satisfied by a distance, because a unitary transformation
represents a rotation in the Hilbert space and the distance
between two states should be invariant under a rotation of the
states.
\item ``Generalized $H$-theorem''. For any complete positive, trace
preserving map $\Phi$ given by
\begin{equation}
\Phi\sigma =\sum_i V_i \sigma V_i^\dag \;\; and \;\; \sum_i
V_i^\dag V_i =1, \label{cp}
\end{equation}
\begin{equation}
S(\Phi\rho\|\Phi\sigma)\leq S(\rho\|\sigma) \label{cp1}
\end{equation}
This is a very significant result because the most general way
that an open quantum system evolves is mathematically represented
by a map of the type given by Eq. (\ref{cp}). Technically, this
kind of map is known as a $CP-map$. Therefore the meaning of Eq.
(\ref{cp1}) is that non-unitary evolution decreases
distinguishability between states. Of course, an unitary evolution
is a particular case of a $CP-map$.

Another example of a $CP-map$ is given by $\sum_i P_i \sigma P_i$
with $P_i$ being a complete set of orthogonal projectors
($P_i^\dagger = P_i$ and $P_i^2 = P_i$). Therefore,
\begin{equation}
S(P_i \rho P_i\|P_i \sigma P_i)\leq S(\rho\|\sigma)
\end{equation}
\item $S$ is jointly convex:
\begin{equation}
S(\sum_i \lambda_i \rho^{(i)}\|\sum_i \lambda_i \sigma^{(i)}) \leq
\sum_i \lambda_i S(\rho^{(i)}\|\sigma^{(i)})
\end{equation}
where the $\lambda_i$ are positive real numbers such that $\sum_i
\lambda_i =1$.
\item Let $\rho^{AB}$ and $\sigma^{AB}$ be two density matrices
of a composite system $AB$ (represented by the Hilbert space
${\cal{H}}_A \otimes {\cal{H}}_B$). Then
\begin{eqnarray}
S(\rho^A \|\sigma^A) &\leq & S(\rho^{AB} \| \sigma^{AB})
\nonumber \\
S(\rho^B \|\sigma^B) &\leq & S(\rho^{AB} \| \sigma^{AB})
\end{eqnarray}
with $\rho^A = Tr_B \rho^{AB}$ and $\rho^B = Tr_A \rho^{AB}$ (here
$Tr_A$ and $Tr_B$ represent the partial trace operators). These
inequalities have a very natural interpretation: to take the trace
over a part of a system leads to a loss of information and
therefore it becomes more difficult to distinguish between two
states of the composite system.
\item $S$ is additive in the sense that
\begin{equation}
S(\rho_1\otimes\rho_2\|\sigma_1\otimes \sigma_2) =
S(\rho_1\|\sigma_1)+ S(\rho_2\|\sigma_2)
\end{equation}
with $\rho_1, \sigma_1 \epsilon {\cal{S}}_A$ and $\rho_2, \sigma_2
\epsilon {\cal{S}}_B$.
\item The relative entropy verifies Donald's identity \cite{Donald}: Let us suppose
that the density operators $\rho_i$ occur with probability $p_i$,
yielding an average state $\rho = \sum_i p_i \rho_i$ and let
$\sigma$ be an arbitrary density operator. Then
\begin{equation}
\sum_i p_i S(\rho_i\|\sigma)= \sum_i p_i S(\rho_i \| \rho) +
S(\rho\|\sigma)
\end{equation}

\end{enumerate}

\section{An alternative to the relative entropy}
We define the quantum Jensen-Shannon divergence (QJSD) as
\begin{equation}
JS(\rho\|\sigma) = \frac{1}{2}[S(\rho\|\frac{\rho+\sigma}{2})+
S(\sigma\|\frac{\rho+\sigma}{2})] \label{def1}
\end{equation}
which can be rewritten in terms of the von Neumann entropy,
$H_N(\rho) = - \tr(\rho \log \rho)$, as
\begin{equation}
JS(\rho\|\sigma) = H_N(\frac{\rho + \sigma}{2})-\frac{1}{2}
H_N(\rho) - \frac{1}{2} H_N(\sigma)
\end{equation}

If $\rho$ and $\sigma$ are density operators with complete sets of
eigenvectors $\{|r_i\rangle\}$ and $\{|s_i\rangle\}$, such that
$\rho = \sum_i r_i |r_i\rangle\langle r_i|$ and $\sigma = \sum_j
s_j |s_j\rangle\langle s_j|$, then the QJSD between $\rho$ and
$\sigma$ is expressed as
\begin{equation}
JS(\rho\|\sigma)= \frac{1}{2}\{ \sum_{k,i} |\langle t_k|
r_i\rangle|^2 r_i \log(\frac{2 r_i}{\lambda_k}) + \sum_{k,j}
|\langle t_k| s_j\rangle|^2s_j \log(\frac{2 s_j}{\lambda_k}) \}
\end{equation}
with $\lambda_k=\sum_i r_i|\langle t_k|r_i\rangle|^2 + \sum_j s_j
|\langle t_k|s_j\rangle|^2$ and $\{|t_k\rangle\}$ a complete set
of normalized eigenvectors of $\rho+\sigma$.

This quantity is positive, null if and only if $\rho = \sigma$,
symmetric and always well defined. In fact, the restriction
imposed on the supports of $\rho$ and $\sigma$ for the relative
entropy is lifted for the QJSD.

If $\rho$ and $\sigma$ commute they are diagonal in the same
basis, that is,
\begin{equation}
\rho = \sum_i r_i |i \rangle\langle i| \;\;\; \sigma=\sum_i s_i
|i\rangle\langle i| \nonumber
\end{equation}
with $|i\rangle$ an orthonormal basis. Then
\begin{equation}
JS(\rho\|\sigma) = JS(\{r_i\},\{s_i\})
\end{equation}

The QJSD is bounded. In fact, as it is known, the von Neumann
entropy satisfies the following inequality \cite{Nielsen}: if
$\rho = \sum_i p_i \rho_i $ is a mixture of quantum states
$\rho_i$ with $p_i$ a set of positive real numbers such that
$\sum_i p_i =1$, then
\[
H_N(\sum_i p_i \rho_i) \leq \sum_i p_i H_N(\rho_i) + H_S(\{p_i\})
\]
The equality is attained if and only if the states $\rho_i$ have
support on orthogonal subspaces.  By putting $p_1 = p_2 =
\frac{1}{2}$ and $\rho_1 = \rho$ and $\rho_2 = \sigma$ in the
inequality above we have
\begin{equation}
0 \leq JS(\rho \| \sigma) \leq 1 \label{bound}
\end{equation}

Properties 1 to 4 of relative entropy are inherited by the QJSD
and their validity can be checked directly from the representation
(\ref{def1}). For example, property 2 for the QJSD can be proved
as follows:
\begin{eqnarray}
JS(\Phi \rho\|\Phi \sigma) &=& \frac{1}{2}[S(\Phi\rho\|\frac{\Phi
\rho +\Phi \sigma}{2}) + S(\Phi \sigma\| \frac{\Phi \rho + \Phi
\sigma}{2})] \nonumber
\\
&=& \frac{1}{2}[S(\Phi \rho\| \frac{\Phi (\rho + \sigma)}{2})
+ S(\Phi \sigma| \frac{\Phi (\rho + \sigma)}{2})] \nonumber \\
& \leq & \frac{1}{2}[S(\rho\| \frac{\rho + \sigma}{2}) + S(\sigma
\| \frac{\rho + \sigma}{2})] = JS(\rho\|\sigma)
\end{eqnarray}

Additivity does not remain valid for the QJSD. However, the QJSD
verifies a ``restricted additivity'':
\begin{equation}
JS(\rho_1\otimes\rho_2\|\sigma_1\otimes\rho_2) =
JS(\rho_1\|\sigma_1) \label{ra}
\end{equation}
with $\rho_1, \sigma_1 \epsilon {\cal{S}}_A$ and $\rho_2 \epsilon
{\cal{S}}_B$. This is an important point that deserves some
attention. We claim that (\ref{ra}) is enough to prove property 4
for the QJSD. To prove this, we follow the same steps as Nielsen
and Chuang in their book where the relative entropy is studied
\cite{Nielsen}. It can be shown that there are unitary
transformations $U_j$ on the space corresponding to the part $B$
and numbers $\lambda_j, (\sum_j \lambda_j=1)$ such that
\[
\rho^A \otimes \frac{I}{d} = \sum_j \lambda_j U_j \rho^{AB}
U^\dagger_j
\]
for all density operator $\rho^{AB}$ of the composite system $AB$.
Here $d$ is the dimension of the Hilbert space of part $B$ and $I$
its identity operator. Hence, from convexity, the invariance under
unitary evolution and the restricted additivity verified by the
QJSD, we have:
\begin{eqnarray}
JS(\rho^A\|\sigma^A) & = & JS(\rho^A \otimes \frac{I}{d}\|
\sigma^A \otimes \frac{I}{d}) \nonumber \\
& \leq & \sum_j \lambda_j JS(U_j \rho^{AB}U^\dagger_j\|U_j
\sigma^{AB} U^\dagger_j) \nonumber \\
& \leq & \sum_j \lambda_j JS(\rho^{AB}\|\sigma^{AB}) =
JS(\rho^{AB}\|\sigma^{AB})
\end{eqnarray}
Of course, due to the linearity of the partial trace operator,
this property can be derived from the analogous ones satisfied by
the relative entropy.

The non-increasing character of the QJSD upon a $CP-map$ has the
following consequence: Let $\{E_{\alpha}\}$ be a set of positive -
operator - valued measure (POVM), that is, a set of Hermitian
operators $E_{\alpha} \geq 0$ that satisfy $\sum_{\alpha}
E_{\alpha} = 1$. It is known that every generalized quantum
measurement can be described by a POVM \cite{Nielsen}. Given two
operators $\rho$ and $\sigma$ belonging to $\cal{S}$, we can
introduce the probability distributions
\[
p_{\alpha} = \tr(\rho E_{\alpha}) \;\;\; q_{\alpha} = \tr(\sigma
E_{\alpha})
\]
Following  Vedral \cite{Vedral} we define
\[
JS_1(\rho\|\sigma) = \sup_{\{E_{\alpha}\}}JS(\{p_{\alpha}\},
\{q_{\alpha}\})
\]
where the supremum is taken over all POVM's. Given that
$S(\rho\|\sigma) \geq S_1(\rho\|\sigma)$ with $S_1(\rho\|\sigma) =
\sup_{\{E_{\alpha}\}} S(\{p_{\alpha}\}, \{q_{\alpha}\})$, we can
derive the inequality
\begin{equation}
JS(\rho\|\sigma) \geq JS_1(\{p_{\alpha}\}, \{q_{\alpha}\})
\end{equation}
Of course the map $\rho\rightarrow \tr(\rho E_{\alpha})$ is a
$CP-map$.

To conclude the listing of properties satisfied by the QJSD we
note that starting from Donald's identity, we can obtain a useful
representation for the QJSD:
\begin{equation}
2 JS(\rho\|\sigma)= S(\rho\|\tau) + S(\sigma\|\tau) - 2
S(\frac{\rho + \sigma}{2}\|\tau)
\end{equation}
where $\tau$ is an arbitrary density operator. This identity
allows us to evaluate the QJSD between two density operators in
terms of a third arbitrary one.

In the following discussion we investigate the relation between
the QJSD and other quantum distances. In reference \cite{nosotros}
we showed that the JSD is a good approximation to the Wootters
distance (up to third order in their power expansion). To search
for a similar relationship between the QJSD distance and that
introduced by Braunstein and Caves, let us consider two
neighboring density operators, $\rho$ and $\rho + d\rho$. The QJSD
between them reduces to (up to second order in $\varepsilon$)
\begin{equation}
JS(\rho \|\rho+d\rho)= \frac{\varepsilon^2}{8} \left[\sum_j
\frac{\omega_{jj}}{p_j}+\sum_{j,k} (p_j-p_k)(\log p_j - \log p_k)
|a_{kj}|^2 \right] \label{app}
\end{equation}
where we have assumed that $d\rho \equiv \varepsilon\omega$, with
$ \varepsilon \ll 1$. The $\omega_{jk}$ are the matrix elements of
$\omega$ in the basis that diagonalizes $\rho$, that is $\rho =
\sum_j p_j |j \rangle \langle j|$, and the eigenvectors of $\rho +
d\rho$ are written (up to first order in $\varepsilon$) in the
form:
\[
 |j^{(1)} \rangle = \sum_k (\delta_{kj} + \varepsilon a_{jk}) |k\rangle
\]
By expanding $\log p_j -\log p_k = \frac{2(p_j - p_k)}{p_j + p_k}
- \frac{4(p_j - p_k)^3}{(p_j + p_k)^3} + ...$ and after some
algebra, the expression (\ref{app}) can be rewritten as
\begin{equation}
JS(\rho \|\rho+d\rho)= \frac{\varepsilon^2}{8} \left[\sum_j
\frac{\omega_{jj}}{p_j}+ 2 \sum_{j,k} \frac{(p_j-p_k)^2}{p_k+p_j}
|a_{kj}|^2 \right]+ \vartheta
\end{equation}
with $\vartheta$ a term that involves sums of powers of $(p_j -
p_k)$ of even order not lower than four. The two first terms
coincide with the metric introduced by Braunstein and Caves  as a
generalization of the Wootters's distance, $ds_{DO}$ \cite{Caves}:
\[
JS(\rho\|\rho+d\rho) \approx \frac{1}{8} ds^2_{DO}
\]
As these authors further note, the metric $ds_{DO}$ is related to
the Bures' metric, $B(\rho\|\sigma) = \surd 2
[(1-\tr(\rho^{1/2}\sigma \rho^{1/2})^{1/2}]^{1/2}$:
\[
4B^2(\rho\|\rho+d\rho)) \approx ds^2_{DO}
\]
Therefore, we can conclude that for two neighboring states
\begin{equation}
B(\rho\|\rho+d\rho) \approx \sqrt{2 JS(\rho\|\rho+d\rho)}
\label{Bures}
\end{equation}

To finish this section, we proceed to evaluate the QJSD in a
particular case. Let us consider the density operator $\rho_W$
corresponding to a Werner state, for a system of two 1/2-spin
particles \cite{Werner}:
\[
\rho_W = F|\Psi^-\rangle\langle\Psi^-| + \frac{1-F}{3}
\left(|\Psi^+\rangle\langle\Psi^+|+|\Phi^+\rangle
\langle\Phi^+|+|\Phi^-\rangle\langle\Phi^-| \right)
\]
with $|\Psi^{\pm}\rangle^= \frac{1}{\surd
2}\left(|\uparrow\downarrow\rangle\pm|\downarrow\uparrow\rangle
\right)$ and $|\Phi^{\pm}\rangle^=\frac{1}{\surd
2}(|\uparrow\uparrow\rangle \pm|\downarrow\downarrow\rangle)$. $F$
is the purity of the state $\rho_W$ with respect to the reference
state $\sigma=|\Psi^-\rangle\langle\Psi^-|$. Then, in terms of the
purity $F$, the QJSD between $\rho_W$ and $\sigma$, results:
\begin{equation}
JS(\rho_W\|\sigma)= \frac{1}{2}[F\log F - (1+F)
\log(\frac{1+F}{2})]
\end{equation}
The fact that we can evaluate the $JS(\rho\|\sigma)$ when one of
its arguments is a pure state, is a clear advantage of the QJSD
over the relative entropy, which becomes divergent in this case.

\section{Discussion}
In this section we suggest possible applications of the QJSD in
the context of quantum information theory. Here we show that the
proposed distance is useful to generate an entanglement measure,
as well as an alternative concept of fidelity. We also give an
interesting interpretation of QJSD as the upper bound for the
accessible quantum information. A more detailed study of each one
of these proposals will be presented in a future work.

\subsection{Entanglement}
Let $\cal{H} = \cal{H}_A\otimes \cal{H}_B$ be the Hilbert space of
a quantum system consisting of two subsystems $A$ and $B$.
According to Vedral et. al. any measure of entanglement $\cal{E}$
has to satisfy the following (necessary) conditions
\cite{Vedral2}:
\begin{itemize}
\item i) ${\cal{E}}(\rho)=0 \;\; iff\;\; \rho$ is separable. Let us
recall that a state is separable if it can be expressed as
$\rho=\sum_j \lambda_j \rho^{(j)}_1 \otimes \rho^{(j)}_2$ for some
density operators $\rho^{(j)}_1 \epsilon \cal{S}_A$ and
$\rho^{(j)}_2 \epsilon \cal{S}_B$, and $\sum_j \lambda_j =1, \;
\lambda_j \geq 0$.
\item ii) ${\cal{E}}(\rho) = {\cal{E}}(U_A \otimes U_B \rho U_A^\dag \otimes
U_B^\dag)$ for two arbitrary unitary operators $U_A$ and $U_B$
acting on the corresponding Hilbert space.
\item iii) ${\cal{E}}(\rho)$ is convex with respect to its
argument:
\[
{\cal{E}}(\sum_j \lambda_j \rho_j) \leq \sum_j \lambda_j
{\cal{E}}(\rho_j)
\]
\item iv) ${\cal{E}}(\rho)$ decreases under generic quantum
operations, that is, if $\{V_i\}$ are bounded operators such that
$\sum V_i V_i^\dag = 1$, $ \rho_j = \frac{V_j \rho
V_j^\dag}{\lambda_j}$ and $\lambda_j = \tr(V_j \rho V_j^\dag)$,
then
\[
{\cal{E}}(\rho) \geq \sum_j \lambda_j {\cal{E}}(\rho_j)
\]
\end{itemize}
Vedral and collaborators \cite{Vedral2} showed that an
entanglement measure defined as
\begin{equation}
{\cal{E}}(\rho) = \inf_{\sigma} D(\rho\|\sigma) \label{entan}
\end{equation}
satisfies conditions i)-iv), whenever the distance $D: \cal{S}
\otimes \cal{S} \rightarrow \Re$,  verifies the properties a)-b)
listed below:
\begin{itemize}
\item a) $D(\rho\|\sigma) \geq 0 $ and $D(\rho\|\rho)= 0$ for any
$\rho, \sigma \epsilon \cal{S}$
\item b) $D(\Phi\rho\|\Phi\sigma) \leq D(\rho\|\sigma)$ for any
$CP-map, \Phi$.
\end{itemize}
In (\ref{entan}) the infimum is taken over the set of separable
states.

As we showed in the preceding section, the QJSD $JS(\rho\|\sigma)$
satisfies these two requirements. So we can introduce a new
entanglement measure by
\begin{equation}
{\cal{E}}_{JS}(\rho) = \inf_{\sigma} JS(\rho\|\sigma)
\end{equation}
where, once again, the infimum is taken over all separable states.

\subsection{Fidelity}
The \textit{fidelity} of states $\rho$ and $\sigma$ is defined as
\cite{Schu},\cite{Bennett}
\begin{equation}
F(\rho\|\sigma) = \left( \tr \sqrt{\sqrt{\rho} \sigma \sqrt{\rho}}
\right)^2 \label{fid}
\end{equation}
This quantity is symmetric, invariant under unitary
transformations and bounded between 0 and 1. For a pure state
$\sigma = |\Psi\rangle\langle\Psi|$, fidelity reduces to the
amplitude $\langle\Psi|\rho|\Psi\rangle$.

The fidelity is related to the Bures's metric by the expression
\begin{equation}
B^2(\rho\|\sigma) = 2 \left( 1- \sqrt{F(\rho\|\sigma)} \right)
\end{equation}
Taking Eq. (\ref{Bures}) into consideration we can introduce an
alternative definition of fidelity (\ref{fid}):
\begin{equation}
F_{JS}(\rho\|\sigma) \equiv \left(1- JS(\rho\|\sigma)\right)^2
\label{nfid}
\end{equation}
>From inequalities (\ref{bound}), $F_{JS}$ must be bounded between
0 and 1. In particular, for $\rho = \sigma, \; F_{JS} = 1$ and
$F_{JS}(\rho\|\sigma) = 0$ if and only if $\rho$ and $\sigma$ have
support on orthogonal subspaces.

The proposal (\ref{nfid}) has another justification. As a textbook
exercise it can be shown that the fidelity of states $\rho$ and
$\sigma$ can be evaluated in terms of \textit{purifications} of
these states as
\begin{equation}
F(\rho\|\sigma) = \max_{|\varphi\rangle}
|\langle\psi|\varphi\rangle|^2
\end{equation}
where $|\psi\rangle$ is any fixed purification of $\rho$, and the
maximization is over all the purifications of $\sigma$
\cite{Nielsen}. At this point, it should be recalled that the
Wootters' distance between the two \textit{pure states}
$|\psi\rangle$ and $|\varphi\rangle$ is given by \cite{Wootters}:
\[
W(|\psi\rangle \| |\varphi\rangle) = \arccos
|\langle\psi|\varphi\rangle|
\]
Then, fidelity can be expressed in terms of the Wootters' distance
as
\begin{eqnarray}
F(\rho\|\sigma) & = & \max_{|\varphi\rangle} \cos^2
(W(|\psi\rangle \| |\varphi\rangle)) \nonumber \\
& = & \max_{|\varphi\rangle} (1- W^2(|\psi\rangle \|
|\varphi\rangle))+ ...)
\end{eqnarray}
Now taking into account the relation between the Wootters's
distance and the JSD divergence \cite{nosotros}
\[
W(|\psi\rangle \| |\varphi\rangle) \approx \sqrt{2 JS(|\psi\rangle
\| |\varphi\rangle)},
\]
we can approximate
\begin{equation}
F(\rho\|\sigma) \approx \max_{|\varphi\rangle} \left(1 -
JS(|\psi\rangle \| |\varphi\rangle) \right)^2
\end{equation}
which, we think, might be taken as an interesting starting point
to investigate (\ref{nfid}) as a fidelity measure.

\subsection{Quantum information accessibility}
Holevo \cite{Holevo} proved that, if a system is prepared in a
state $X$ chosen from one of the density operators $\rho_i,
(i=1,...,n)$ with probability $p_i$, then the (mutual) information
$I(X:Y)$ that can be gathered about the identity of the state $X$
by a POVM measurement, with outcome $Y$, is bounded according to
the inequality
\begin{equation}
I(X:Y) \leq \chi = H_N(\sum_i p_i \rho_i) - \sum_i p_i H_N(\rho_i)
\end{equation}
The quantity $\chi$ is precisely the extension of expression
(\ref{varios}) to density operators with the weights $\pi_i$
replaced by the probabilities $p_i$, that is, $\chi\equiv
JS^{(p_1,...p_n)}(\rho_1\|...\|\rho_n)$. This fact provides an
interesting interpretation of the generalized QJSD.

\begin{center}
AKNOWLEDGMENT
\end{center}
We are grateful to Secretaria de Ciencia y Tecnica de la
Universidad Nacional de C\'ordoba for financial assistance. This
work was partially supported by Grant BIO2002-04014-C03-03 from
the Spanish Government.

\end{document}